\documentclass[aps,prb,reprint,superscriptaddress]{revtex4-2}
\usepackage{amsmath}
\usepackage{amssymb}
\usepackage{graphicx}
\usepackage{hyperref}
\usepackage{color,xcolor}
\begin{document}
\title{Manipulation of $J_{\rm eff}=3/2$ states by tuning the tetragonal distortion}
\author{Yakui Weng}
\email{Email: wyk@njupt.edu.cn}
\affiliation{School of Science, Nanjing University of Posts and Telecommunications, Nanjing 210023, China}
\author{Shuai Dong}
\email{Email: sdong@seu.edu.cn}
\affiliation{School of Physics, Southeast University, Nanjing 211189, China}
\date{\today}

\begin{abstract}
The spin-orbit entangled quantum states in $4d/5d$ compounds, e.g., the $J_{\rm eff}=1/2$ and $J_{\rm eff}=3/2$ states, have attracted great interests for their unique physical roles in unconventional superconductivity and topological states. Here, the key role of tetragonal distortion is clarified, which determines the ground states of $4d^1$/$5d^1$ systems to be the $J_{\rm eff}=3/2$ one (e.g. K$_2$NbCl$_6$) or $S=1/2$ one (e.g. Rb$_2$NbCl$_6$). By tuning the tetragonal distortion via epitaxial strain, the occupation weights of $d_{xy}$/$d_{yz}$/$d_{xz}$ orbitals can be subtly modulated, competing with the spin-orbit coupling. Consequently, quantum phase transitions between $S=1/2$ state and $J_{\rm eff}=3/2$ state, as well as between different $J_{\rm eff}=3/2$ states, can be achieved, resulting in significant changes of local magnetic moments. Our prediction points out a reliable route to engineer new functionality of $J_{\rm eff}$ states in these quantum materials.
\end{abstract}
\maketitle

\section{Introduction}
Controlling electronic states in materials is one of the most important topics of condensed matter physics, which can lead to new functional devices. In strongly correlated electronic systems, the interactions among multiple degrees of freedom (spin, lattice, orbital, and charge) establish subtle balances among rich quantum states, which provide unique opportunities for abundant functionalities, including but not limited to high-temperature superconductivity, colossal magnetoresistivity, and multiferroicity \cite{Imada:Rmp,Dagotto:Sci,Dagotto:Mrs,Chakhalian:Rmp14,Zubko:Arcmp11,Dong:Ap,Dong:Nsr19}.

In the past decade, the spin-orbit coupling (SOC) has attracted more and more attentions in many branches of condensed matter. In particular, for those heavy transition metal elements with $4d$ or $5d$ orbitals, the synactic effect of SOC and electron correlation can lead to emergent quantum phenomena such as topological phase \cite{Pesin:Np10,Wan:Prb11,Krempa:Prb12}, unconventional superconductivity \cite{Wang:Prl11,Watanabe:Prl13,Yan:Prx15,Kim:Np16}, Kitaev spin liquid \cite{Jackeli:Prl09,Takagi:Nrp19}, large anisotropic magnetoresistivity \cite{Lu:Afm18,Wang:Nc19}, as well as magnetic quadrupole moments \cite{Weng:Prb20,Ishikawa:Prb19}. These SOC-entangled quantum states should be highly sensitive to external stimulations, which provides the opportunities to manipulate them. However, in such an emergent field, most attentions have been paid on the realization of these quantum states and their unique physical properties, while the manipulation of these states are rarely touched till now.

Our recent theoretical work predicted that the $4d^1$ electron in K$_2$NbCl$_6$ is in the $J_{\rm eff}=3/2$ state, which could be tuned to the $S=1/2$ state by rotating the spin axis \cite{Weng:Prb20}. A key characteristic of the ideal $J_{\rm eff}=3/2$ state is that its spin magnetic moment and orbital magnetic moment should cancel each other, leading to a quenched net magnetic dipole moment (i.e., ideally $0$ $\mu_{\rm B}$/Nb). However, a recent neutron study found that its sister compound Rb$_2$NbCl$_6$ exhibited a large magnetic moment up to $0.96$ $\mu_{\rm B}$/Nb \cite{Ishikawa:Prb19}, which was close to the $S=1/2$ limit instead. For comparison, $A_2$TaCl$_6$ ($A$=K, Rb, Cs) with $5d^1$ electronic configuration exhibited much smaller magnetic moments $\sim0.25$-$0.30$ $\mu_{\rm B}$/Ta \cite{Ishikawa:Prb19}, closer to the $J_{\rm eff}=3/2$ case, although not so ideal.

In this work, the underlying physics of these $J_{\rm eff}=3/2$ candidates will be studied theoretically, to clarify why some of them are the $S=1/2$ state while some are close to the $J_{\rm eff}=3/2$ one. Besides the previously known SOC and Hubbard $U$, the tetragonal distortion (i.e., the static Jahn-Teller $Q_3$ mode) is found to play a decisive role in $J_{\rm eff}=3/2$ states of $4d^1/5d^1$ systems. Furthermore, quantum phase transitions between the $S=1/2$ and $J_{\rm eff}=3/2$ states, as well as between different $J_{\rm eff}=3/2$ states, are tuned by epitaxial strain. Although the influences of Jahn-Teller distortion on physical properties have been touched more or less in these SOC systems  \cite{Xu:Npjqm16,Iwahara:Prb18,Mosca:Prb21,Liu:Prl19,Plotnikova:Prl16,Henke:Zk,Ishikawa:Prb19,Streltsov:Prx20}, its essential role to the $J_{\rm eff}$ states, especially how to manipulate the $J_{\rm eff}$ states via the strain, has not been reported yet.

\section{Model \& Methods}
The atomic SOC model Hamiltonian can be written as:
\begin{equation}
H_{\rm SOC}=\lambda\textbf{L}\cdot\textbf{S},
\end{equation}
where $\lambda$ is the SOC coefficient; $\textbf{L}$ is the orbital moment operator; and $\textbf{S}$ is the spin operator.

In the octahedral crystal field, the $d$ orbitals will be split into the low-lying $t_{\rm 2g}$ triplets ($d_{xz}$/$d_{yz}$/$d_{xy}$) and higher-energy $e_{\rm g}$ doublets ($d_{3z^2-r^2}$/$d_{x^2-y^2}$). For electron configurations from $d^1$ to $d^3$, the $e_{\rm g}$ doublets can be neglected, and only the $t_{\rm 2g}$ triplets are essential. If the tetragonal distortion is considered, there will be an on-site energy splitting between $d_{xz}$/$d_{yz}$ and $d_{xy}$. The details of the SOC model Hamiltonian for $t_{\rm 2g}$ levels and its solution can be found in Supplemental Materials (SM) \cite{supp3}.

In addition to the model study, concrete materials are also studied by density functional theory (DFT). Our DFT calculations were performed using the projector augmented wave (PAW) pseudopotentials as implemented in the Vienna $ab$ $initio$ Simulation Package (VASP) code \cite{Blochl:Prb2,Kresse:Prb99}. The revised Perdew-Burke-Ernzerhof for solids (PBEsol) functional and the generalized gradient approximation (GGA) method are adopted to describe the crystalline structure and electron correlation \cite{Perdew:Prl08}. Using the Dudarev implementation \cite{Dudarev:Prb}, the Hubbard repulsion $U_{\rm eff}$'s are imposed on Nb's $4d$ and Ta's $5d$ orbitals. The cutoff energy of plane-wave is $450$ eV and the $7\times7\times5$ Monkhorst-Pack \textit{k}-point mesh is centered at $\varGamma$ point. Both the lattice constants and atomic positions are fully relaxed until the Hellman-Feynman forces converged to less than $0.01$ eV/{\AA}.

\section{Results \& Discussion}
\subsection{Atomic orbital model}
\begin{figure}
\centering
\includegraphics[width=0.46\textwidth]{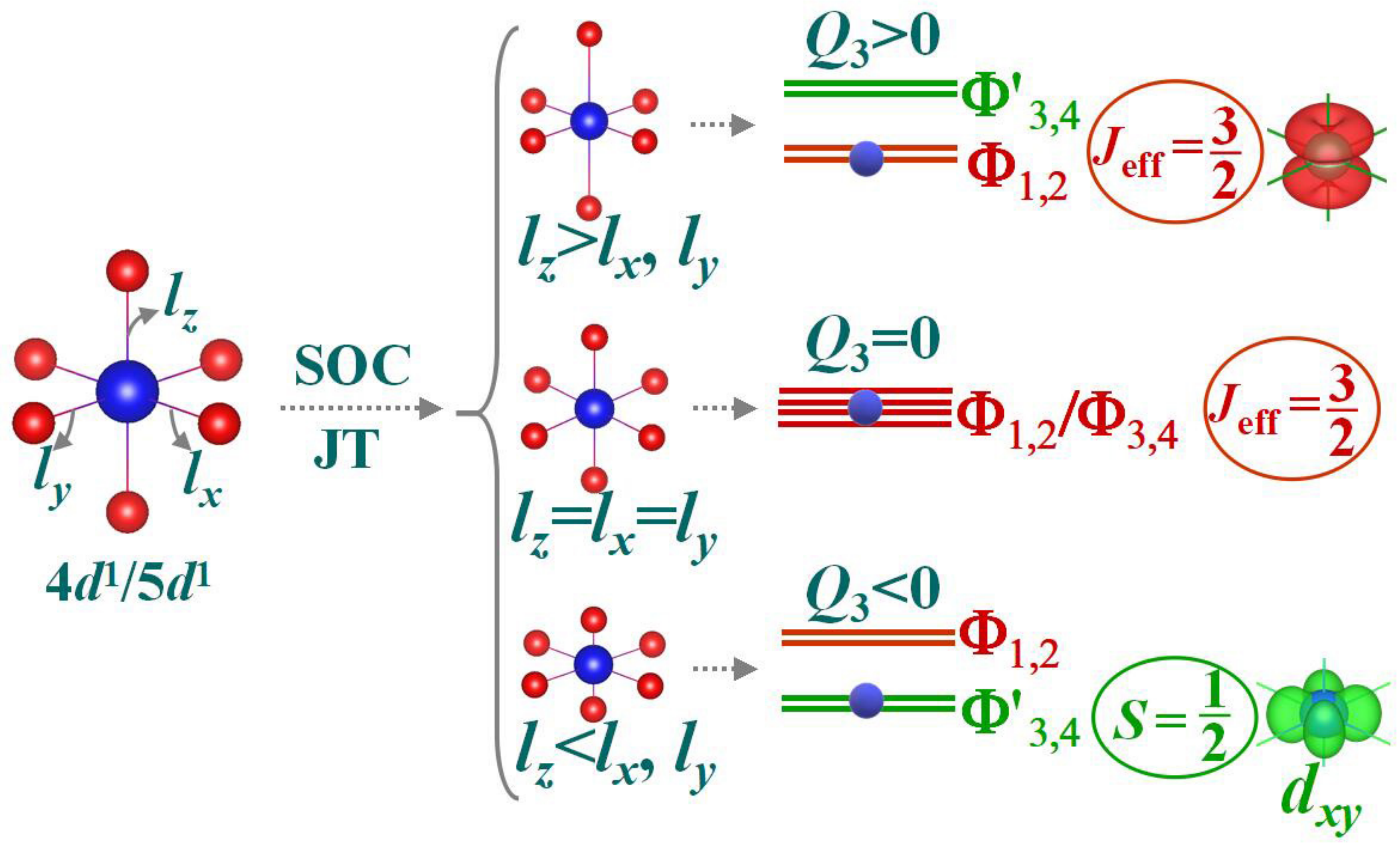}
\caption{Schematic of collaborative effect of SOC and $Q_3$ mode distortion for the $4d^1$/$5d^1$ electron states in the octahedral crystal field. Here, $Q_3$ is defined as $(2l_z-l_x-l_y)/\sqrt{6}$, where $l$ denotes the bond length along a particular axis \cite{Dong:Prb11}. For an undistorted octahedron (i.e., $Q_3=0$), the SOC leads to four degenerate low-lying $J_{\rm eff}=3/2$ levels ($\Phi_1$, $\Phi_2$, $\Phi_3$, and $\Phi_4$) and high-lying $J_{\rm eff}=1/2$ doublets (not shown here). For the $d^1$ configuration, the $J_{\rm eff}=3/2$ levels are quarter-filled. For nonzero $Q_3$, the $J_{\rm eff}=3/2$ levels are split into two groups: $\Phi_1$/$\Phi_2$ and $\Phi_3'$/$\Phi_4'$, the latter of which are mixed with the $J_{\rm eff}=1/2$ states more or less. Once this mixture is heavy, $\Phi_3'$/$\Phi_4'$ are close to the $S=1/2$ states. The sign of $Q_3$ determines whether $\Phi_1$ ($\Phi_2$) or $\Phi_3'$ ($\Phi_4'$) is the lowest energy one.}
\label{F1}
\end{figure}

To give an elegant physical scenario, let's start from a local atomic orbital model. The atomic SOC can split the $t_{\rm 2g}$ levels into the low-lying $J_{\rm eff}=3/2$ quartets and higher-energy $J_{\rm eff}=1/2$ doublets, as shown in Fig.~\ref{F1}. The wave functions of these low-lying $J_{\rm eff}=3/2$ states can be expressed using the $t_{\rm 2g}$ bases ($d_{xy}$, $d_{yz}$, $d_{xz}$) as following \cite{supp3}:
\begin{eqnarray}
\nonumber\Phi_1&=&\frac{1}{\sqrt{2}}(|d_{yz}\uparrow>+i|d_{xz}\uparrow>),\\
\nonumber\Phi_2&=&\frac{1}{\sqrt{2}}(|d_{yz}\downarrow>-i|d_{xz}\downarrow>),\\
\nonumber\Phi_3&=&\frac{1}{\sqrt{6}}(|d_{yz}\downarrow>+i|d_{xz}\downarrow>-2|d_{xy}\uparrow>),\\ \Phi_4&=&\frac{1}{\sqrt{6}}(|d_{yz}\uparrow>-i|d_{xz}\uparrow>+2|d_{xy}\downarrow>),
\end{eqnarray}
where $\uparrow$/$\downarrow$ denote the spin up/down. And the wave functions of higher-energy $J_{\rm eff}=1/2$ states can be expressed as:
\begin{eqnarray}
\nonumber\Phi_5&=&\frac{1}{\sqrt{3}}(|d_{yz}\downarrow>+i|d_{xz}\downarrow>+|d_{xy}\uparrow>),\\
\Phi_6&=&\frac{1}{\sqrt{3}}(|d_{yz}\uparrow>-i|d_{xz}\uparrow>-|d_{xy}\downarrow>).
\end{eqnarray}

Then, for the $d^1$ configuration, the $J_{\rm eff}=3/2$ levels are quarter-filled, as sketched in Fig.~\ref{F1}. The one electron can occupy on any of $\Phi_i$'s ($i$=$1-4$) or their linear combinations. The rest three levels can be pushed upward in energy by the Hubbard repulsion, and a Mott gap can be opened if the Hubbard $U$ is enough, resulting in a $J_{\rm eff}=3/2$ Mott state. Such a scenario is a standard one to demonstrate the $J_{\rm eff}$ states, similar to the $J_{\rm eff}=1/2$ case \cite{Kim:Prl08}.

\begin{figure}
\centering
\includegraphics[width=0.46\textwidth]{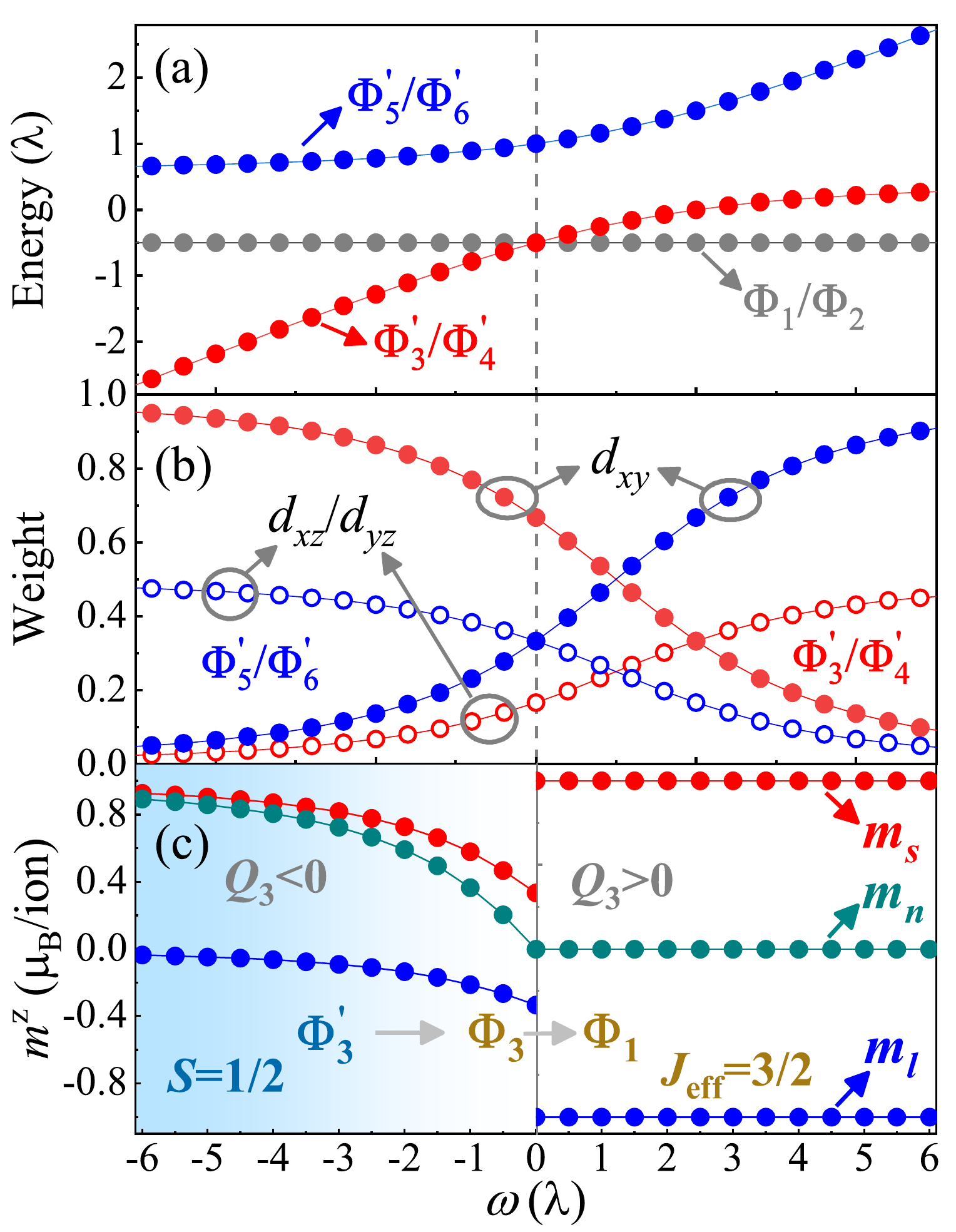}
\caption{Numerical results of atomic orbital model to demonstrate the scenario proposed in Fig.~\ref{F1}. (a) Eigen energies, (b) $t_{\rm 2g}$ orbital weights, and (c) spin ($m_s$) and orbital ($m_l$) magnetic moments (along the $z$-axis), as a function of $Q_3$ intensity, characterized by $\omega$. The $\omega$ and eigen energy are in unit of $\lambda$, the SOC coefficient. In (b), the red and blue symbols are for $\Phi_3'$ ($\Phi_4'$) and $\Phi_5'$ ($\Phi_6'$), respectively. $m_n$: net magnetic moment ($=m_l+m_s$).}
\label{F2}
\end{figure}

However, for such a system with partially occupied degenerate levels, the tetragonal distortion should be active. In fact, in all $A_2$NbCl$_6$ and $A_2$TaCl$_6$ (as well as Sr$_2$IrO$_4$), their octahedra are distorted, with different bond lengths along the $z$-axis and in the $x-y$ plane, as sketched in Fig.~\ref{F1}. Such a tetragonal distortion can lift the degeneracy between $d_{xy}$ and $d_{xz}$/$d_{yz}$, characterized by the coefficient $\omega$ (the ratio of distortion energy splitting and SOC coefficient, as explained in SM \cite{supp3}), which is in proportional to the intensity of $Q_3$ mode distortion. Then the eigen energies [Fig.~\ref{F2}(a)] and corresponding wave functions of $t_{\rm 2g}$ levels can be obtained.

Comparing with the undistorted case, $\Phi_1$ and $\Phi_2$ are not affected since they only involve $d_{xz}$ and $d_{yz}$. However, $\Phi_3'$ and $\Phi_4'$ deviate from their ideal limits, by mixing some $J_{\rm eff}=1/2$ components. In particular, $\Phi_i'$ ($i=3-6$) can be expressed as:
\begin{eqnarray}
\nonumber\Phi_3'&=&\frac{1}{\sqrt{2+a^2}}(|d_{yz}\downarrow>+i|d_{xz}\downarrow>+a|d_{xy}\uparrow>),\\ \nonumber\Phi_4'&=&\frac{1}{\sqrt{2+a^2}}(|d_{yz}\uparrow>-i|d_{xz}\uparrow>-a|d_{xy}\downarrow>),\\
\nonumber\Phi_5'&=&\frac{1}{\sqrt{2+b^2}}(|d_{yz}\downarrow>+i|d_{xz}\downarrow>+b|d_{xy}\uparrow>),\\ \Phi_6'&=&\frac{1}{\sqrt{2+b^2}}(|d_{yz}\uparrow>-i|d_{xz}\uparrow>-b|d_{xy}\downarrow>),
\end{eqnarray}
where $a=[(\omega-1)-\sqrt{(\omega-1)^2+8}]/2$ and $b=[(\omega-1)+\sqrt{(\omega-1)^2+8}]/2$. Then the evolution of $d_{xy}$, $d_{yz}$, $d_{xz}$ weights in $\Phi_i'$ ($i=3-6$) can be obtained  as a function of $\omega$, as shown in Fig.~\ref{F2}(b).

In particular, for the tetragonal elongation ($Q_3>0$, i.e., $\omega>0$), $\Phi_1$ and $\Phi_2$ are lower in energy than $\Phi_3'$ and $\Phi_4'$, since the on-site energy of $d_{xy}$ is higher than that of $d_{xz}$/$d_{yz}$. Thus, the ground state remains the ideal $J_{\rm eff}=3/2$ state. In contrast, for the tetragonal compression ($Q_3<0$), the lower energy $d_{xy}$ orbital makes $\Phi_3'$ and $\Phi_4'$ more favorable. For the ideal $\Phi_3$ or $\Phi_4$, the weight of $d_{xy}$ orbital already reaches $66.7\%$. The negative $Q_3$ ($\omega$) will further enhance its weight. In the large limit of negative $Q_3$, $\Phi_3'$ or $\Phi_4'$ will be eventually close to the pure $d_{xy}$ orbital, i.e., a $S=1/2$ state. In this sense, the tetragonal distortion can tune the phase transitions between the $J_{\rm eff}=3/2$ states and $S=1/2$ one.

The $\omega$-dependent magnetic moments of $\Phi_1$ (or $\Phi_2$) and $\Phi_3'$ (or $\Phi_4'$) are presented in Fig.~\ref{F2}(c), including both the spin component and orbital component. For the positive $Q_3$ case ($\omega>0$), the occupied state is always $\Phi_1$ or $\Phi_2$, which will be split by Hubbard repulsion. The most interesting physical property of $\Phi_1$ (or $\Phi_2$) is that its spin magnetic moment ($1$ $\mu_{\rm B}$) just cancels the orbital magnetic moment ($-1$ $\mu_{\rm B}$), leading to a zero net magnetization \cite{Weng:Prb20}. In the opposite direction, i.e., $\omega<0$, $\Phi_3'$ (or $\Phi_4'$) becomes the low-lying level. With increasing negative $\omega$, $\Phi_3'$'s spin magnetic moment along the $z$-axis gradually increases from $1/3$ $\mu_{\rm B}$ to approach the limit $1$ $\mu_{\rm B}$, while the orbital magnetic moment gradually decreases from $-1/3$ $\mu_{\rm B}$ to the limit $0$ $\mu_{\rm B}$, as expected for the transition from $J_{\rm eff}=3/2$ state to $S=1/2$ state.

\subsection{Materials}
\begin{figure}
\centering
\includegraphics[width=0.46\textwidth]{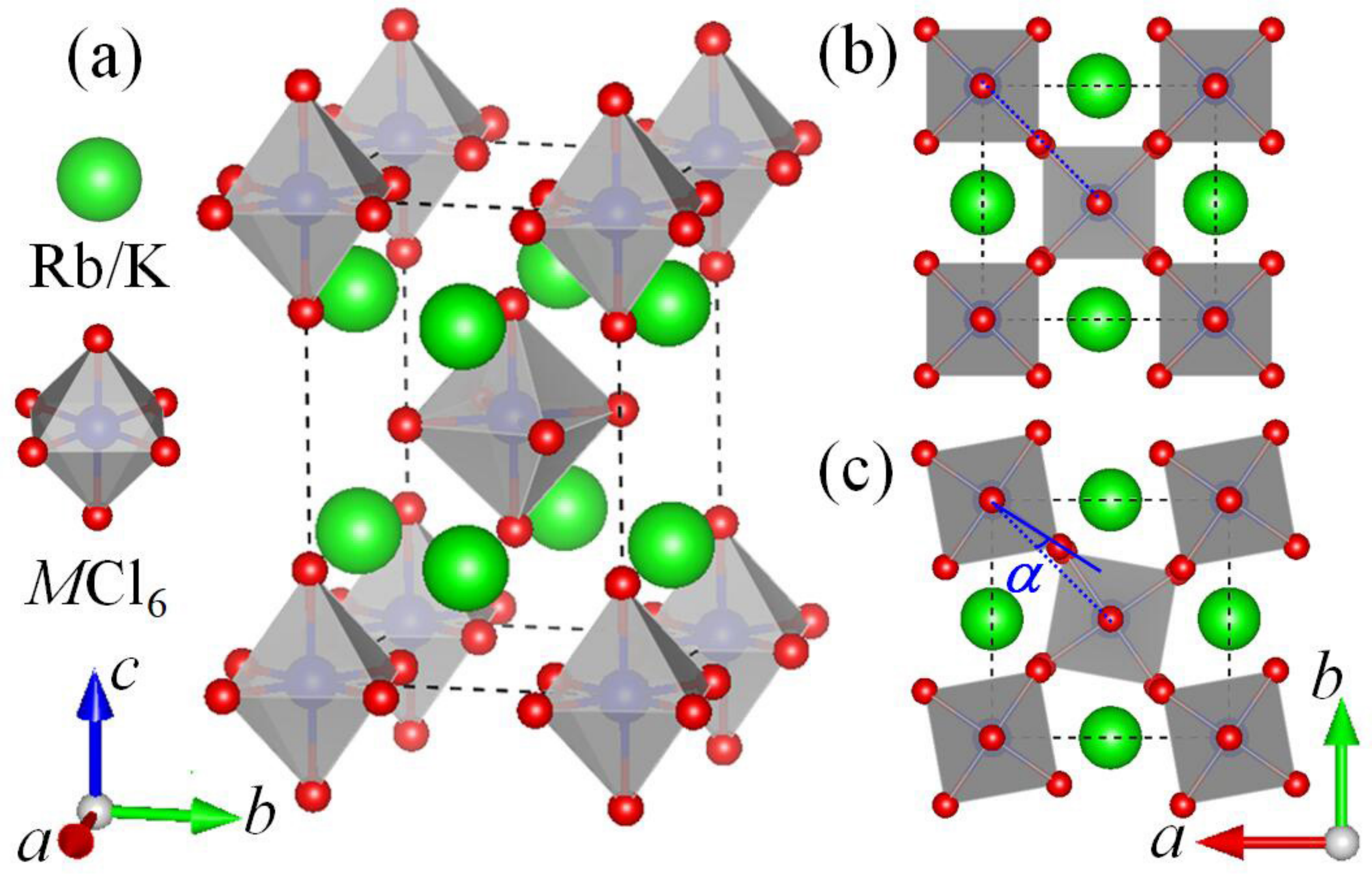}
\includegraphics[width=0.46\textwidth]{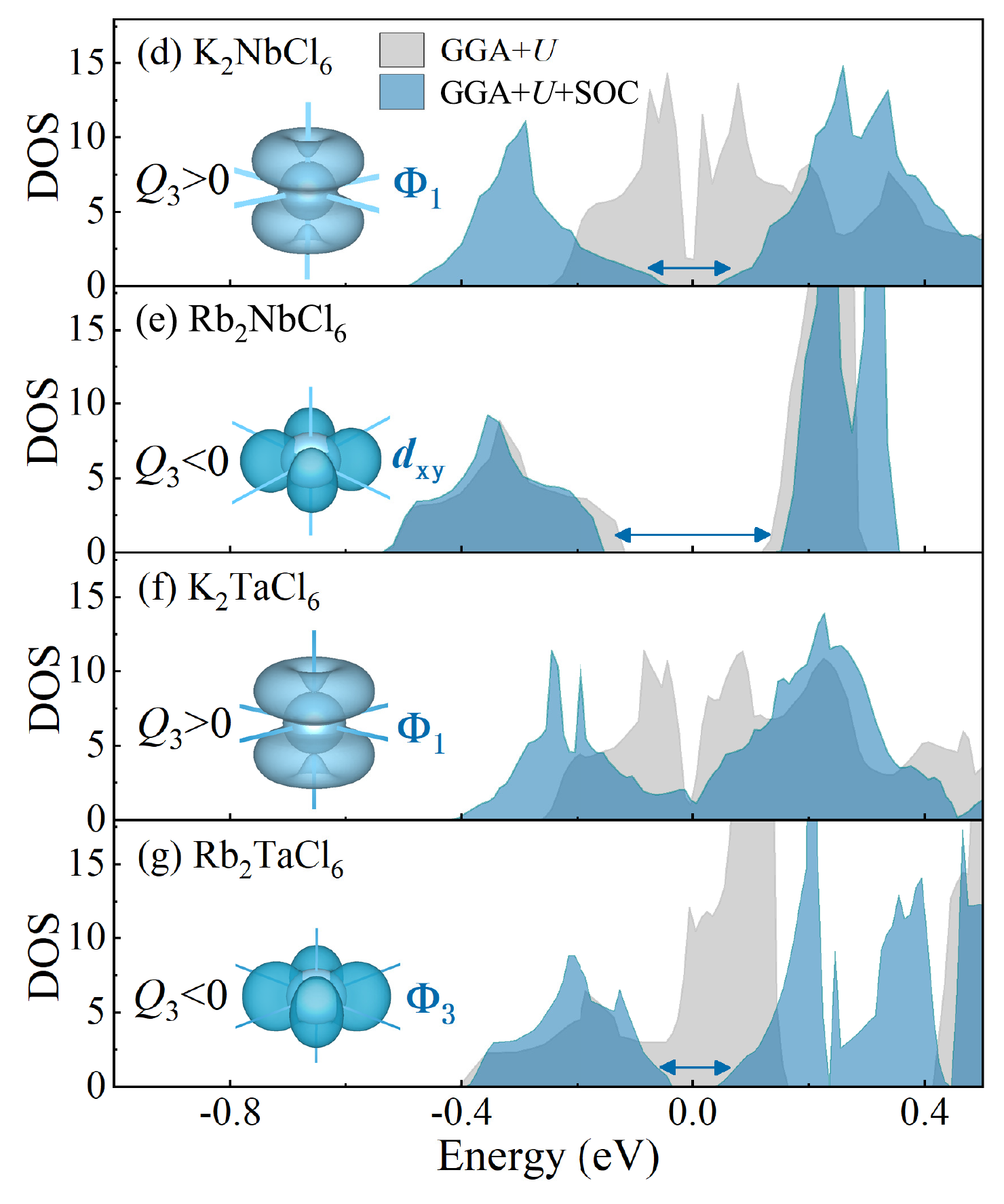}
\caption{(a-c) Schematic crystal structures of $A_2M$Cl$_6$. (b) Top view of $A$=Rb, space group $I4/mmm$ (No. $139$). (c) Top view of $A$=K, space group $P4/mnc$ (No. $128$). The smaller K ion leads to smaller in-plane lattice constant, which results in the elongation (i.e., $Q_3>0$) and rotation of octahedra. Since these octahedra are almost isolated, the effect of octahedral rotation to the electronic states is negligible. (d-g) Comparison of density of states (DOS) between the GGA+$U$ and GGA+$U$+SOC calculations. Here, $U_{\rm eff}=1$ eV and $0.7$ eV are applied on Nb's $4d$ and Ta's $5d$ orbitals, respectively.  Insets: the electron clouds of valence bands near the Fermi level, i.e. the $d^1$ electrons.}
\label{F3}
\end{figure}

The aforementioned physical scenario can be well demonstrated in real materials, e.g. hexachlorides $A_2M$Cl$_6$ ($A$=K, Rb, Cs; $M$=Nb, Ta). The crystal structures of this family are shown in Figs.~\ref{F3}(a-c). Unlike perovskite oxides where oxygen octahedra are connected via the corner-, edge-, or face-sharing manner, these $M$Cl$_6$ octahedra are nearly isolated \cite{Henke:Zk}. In this sense, the electron hoppings between $M$ ions are significantly suppressed, leading to narrow bands near the Fermi level. According to their experimental structures \cite{Ishikawa:Prb19,Henke:Zk}, all $M$Cl$_6$ octahedra in these hexachlorides are tetragonally distorted, i.e., with the static Jahn-Teller $Q_3$ mode distortion. In addition, the $d$-$p$ hybridization in chloride family is much weaker than that in oxides, which makes the $d$ orbital levels more pure. In this sense, these $M$Cl$_6$ octahedra provide an ideal playground to study the SOC-assisted Mottness of its $4d$/$5d$ orbitals.

Previously, some of $A_2M$Cl$_6$ (e.g., $A_2$TaCl$_6$, K$_2$NbCl$_6$) were claimed to host the $J_{\rm eff}=3/2$ state \cite{Ishikawa:Prb19,Weng:Prb20}, while Rb$_2$NbCl$_6$ was found to be close to the $S=1/2$ state \cite{Ishikawa:Prb19}. These puzzling contradictions can be well answered in our DFT calculations \cite{supp3}, as summarized in Figs.~\ref{F3}(d-g), Table~\ref{table1}, as well as Table~S1 in SM \cite{supp3}.

In K$_2$NbCl$_6$, the NbCl$_6$ octahedra are spontaneously elongated along the crystalline $c$-axis (i.e., the $z$-axis of octahedra), i.e., $Q_3>0$, which prefers the pure $J_{\rm eff}=3/2$ state $\Phi_1$ (or $\Phi_2$) [Fig.~\ref{F3}(d)]. The spin and orbital magnetic moments obtained in our DFT calculation are close to their ideal limits (see Table~\ref{table1}). The band gap can only be opened with the help of SOC, implying the SOC-assisted Mottness \cite{Weng:Prb20}.

In contrast, for Rb$_2$NbCl$_6$, the NbCl$_6$ octahedra are spontaneously compressed along the crystalline $c$-axis (i.e., the $z$-axis of octahedra), i.e., $Q_3<0$, which prefers the $\Phi_3'$ (or $\Phi_4'$) state [Fig.~\ref{F3}(e)]. And a band gap is already opened even without SOC, due to the tetragonal distortion splitting. Meanwhile, its spin magnetic moment is large but its orbital magnetic moment is rather small (see Table~\ref{table1}), in agreement with the experimental large magnetic moment \cite{Ishikawa:Prb19}. Therefore, this distorted $\Phi_3'$ (or $\Phi_4'$) is closer to $S=1/2$ ($d_{xy}$), instead of $J_{\rm eff}=3/2$ state $\Phi_3$ (or $\Phi_4$). Previously, the absence of $J_{\rm eff}=3/2$ state in Rb$_2$NbCl$_6$ was attributed to its relative weak SOC intensity of $4d$ orbitals \cite{Ishikawa:Prb19}. However, the present work suggests a more decisive role of $Q_3$ mode distortion in Rb$_2$NbCl$_6$, since its sister member K$_2$NbCl$_6$ can exhibit the pure $J_{\rm eff}=3/2$ state \cite{Weng:Prb20}, even its SOC intensity is not as strong as the $5d$ counterparts.

The story of K$_2$TaCl$_6$ is very similar to K$_2$NbCl$_6$, with a positive $Q_3$ and an even larger SOC, which can lead the $\Phi_1$ (or $\Phi_2$) state, as shown in Fig.~\ref{F3}(f). The only difference is that the weaker Hubbard $U$ of $5d$ orbitals may be not insufficient to open a Mott gap.

The story of Rb$_2$TaCl$_6$ is unique. Its $Q_3$ mode is negative, which seems to be similar to Rb$_2$NbCl$_6$. However, the large SOC of $5d$ orbitals makes a small $\omega$ (the ratio between tetragonal distortion splitting and SOC coefficient). Thus the $5d^1$ state in Rb$_2$TaCl$_6$ is close to $J_{\rm eff}=3/2$ ($\Phi_3$ or $\Phi_4$), instead of $S=1/2$ ($d_{xy}$). As an evidence of $\Phi_3$, its spin and orbital magnetic moments are close to $1/3$ and $-1/3$ $\mu_{\rm B}$/Nb respectively (see Table~\ref{table1}).

\begin{table}
\caption{DFT calculated magnetic moments of $A_2M$Cl$_6$ in unit of $\mu_{\rm B}$/$M$, including the spin magnetic moments ($m_s$), orbital magnetic moments ($m_l$), and the net moments ($m_n=m_s+m_l$).}
\centering
\begin{tabular*}{0.48\textwidth}{@{\extracolsep{\fill}}lcccc}
\hline \hline
	& K$_2$NbCl$_6$ & Rb$_2$NbCl$_6$ & K$_2$TaCl$_6$ & Rb$_2$TaCl$_6$\\
\hline
$m_s$ & $0.877$ & $0.733$ & $0.694$ & $0.343$ \\
$m_l$ & $-0.697$ & $-0.100$ & $-0.522$ & $-0.316$ \\
$m_n$ & $0.180$ & $0.633$ & $0.172$ & $0.027$ \\
\hline
state & $\Phi_1$ & $d_{xy}$ & $\Phi_1$ & $\Phi_3$ \\
\hline \hline
\end{tabular*}
\label{table1}
\end{table}

\subsection{Strain tuning of $J_{\rm eff}=3/2$ states}
Although the $4d^1$ electron in Rb$_2$NbCl$_6$ is not in the $J_{\rm eff}=3/2$ state, it provides an appropriate candidate to study the strain effect. Here the epitaxial biaxial strain is imposed to control the sign and amplitude of $Q_3$ mode. The biaxial strain is defined as $\varepsilon=(a-a_0)/a_0$, where $a_0$ and $a$ are the in-plane lattice constants before and after the strain, respectively. Upon the biaxial strain, the atomic positions and lattice constant long the $c$-axis are further optimized.

\begin{figure}
\centering
\includegraphics[width=0.46\textwidth]{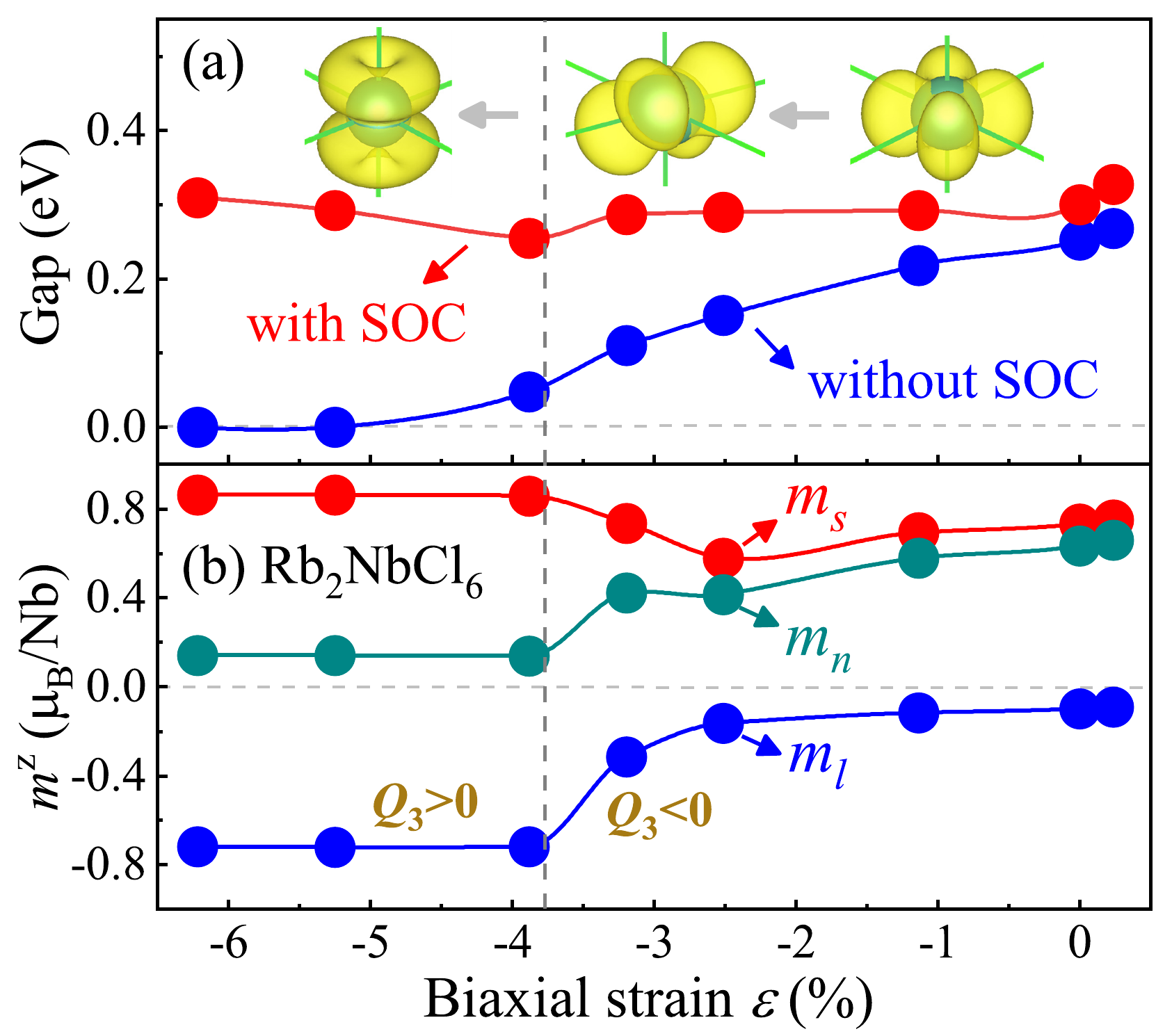}
\includegraphics[width=0.46\textwidth]{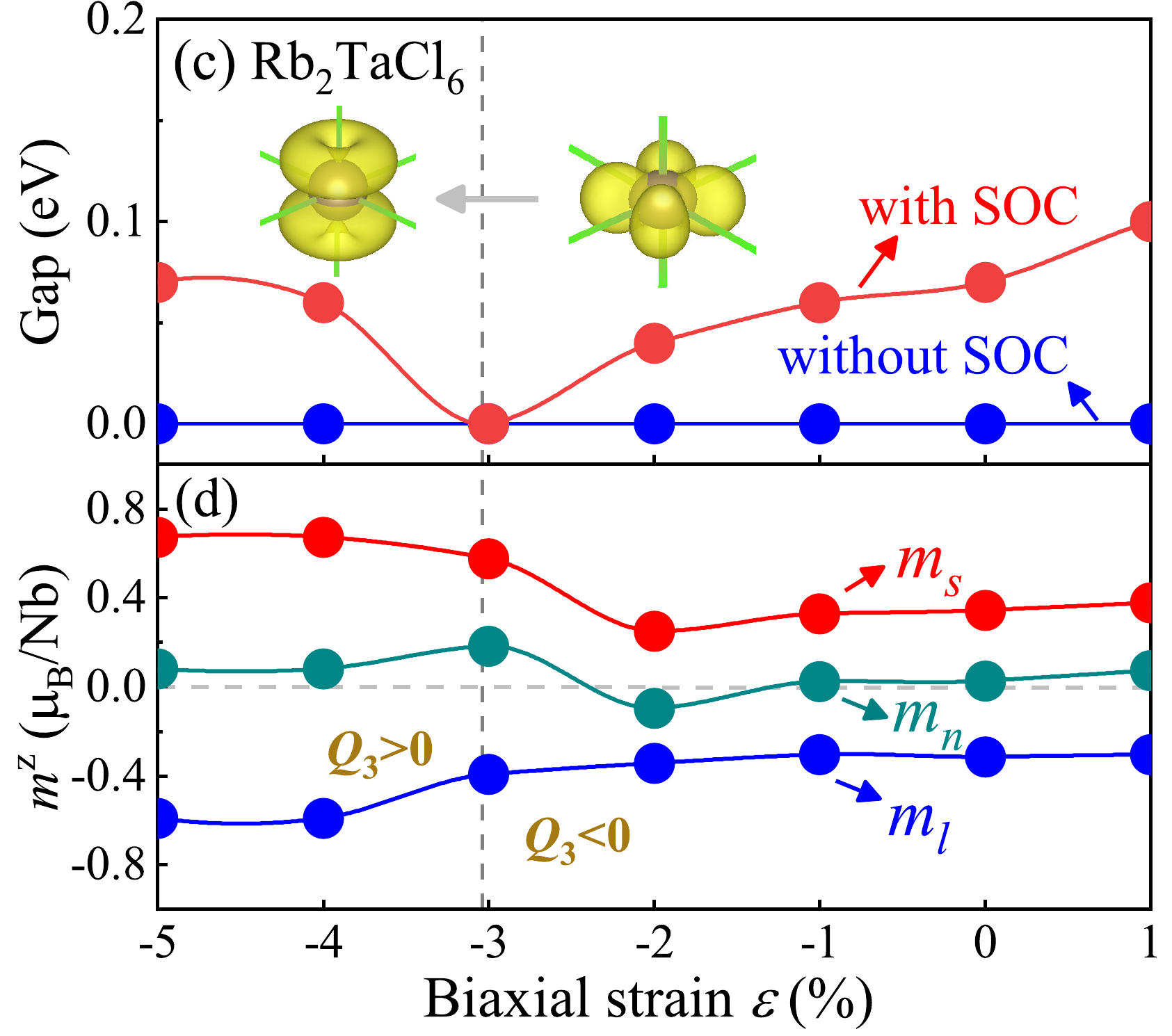}
\caption{Biaxial strain tuned electronic state transitions between $S=1/2$ and $J_{\rm eff}=3/2$ ($\Phi_1$ and $\Phi_3$) in (a-b) Rb$_2$NbCl$_6$ and (c-d) Rb$_2$TaCl$_6$. (a) and (c): The band gaps (with/without SOC). (b) and (d): The magnetic moments of a Nb/Ta ion (with SOC). The boundary $Q_3=0$ (vertical broken line) locates at $\varepsilon\sim-3.8\%$ and $\varepsilon\sim-3.0\%$, respectively. Insets: the electron clouds of valence bands near the Fermi levels.}
\label{F4}
\end{figure}

As shown in Fig.~\ref{F4}(a), without SOC, the band gap is gradually closed with increasing compressive strain. This is due to the half-filling of low-lying degenerate $d_{yz}$ and $d_{xz}$ orbitals in the positive $Q_3$ case. However, the SOC can open the band gap by forming the $J_{\rm eff}$ state ($\Phi_1$ or $\Phi_2$). In other words, the band gap $\sim0.3$ eV in the whole presented region of $\varepsilon$ is driven by different mechanisms: 1) $Q_3$+$U$ in the $S=1/2$ side; 2) SOC+$U$ in the $J_{\rm eff}=3/2$ side.

Correspondingly, the evolution of magnetic moments is shown in Fig.~\ref{F4}(b). When $Q_3$ turns to be positive, the spin moment slightly increases from $\sim0.7$ $\mu_{\rm B}$/Nb to $\sim0.9$ $\mu_{\rm B}$/Nb, while the orbital moment increases rapidly from $\sim-0.1$ $\mu_{\rm B}$/Nb to $\sim-0.7$ $\mu_{\rm B}$/Nb, indicating the transition from the $S=1/2$ state to the $J_{\rm eff}=3/2$ one.

For comparison, the strain effect to Rb$_2$TaCl$_6$ is also studied, which provides an appropriate candidate to tune the two states of $J_{\rm eff}=3/2$ ($\Phi_1$ \textit{vs} $\Phi_3$). As shown in Fig.~\ref{F4}(c), without SOC, the system is always metallic in the whole strain range, due to the weaker Hubbard repulsion ($U_{\rm eff}=0.7$ eV used here) and more spatially expanding wave function of $5d$ orbitals. With increasing compressive strain, the band gap (opened by SOC) gradually decreases to zero and reopened. The gap closing point is just at the $Q_3=0$ boundary, at which $\Phi_i$'s ($i=1-4$) are originally degenerated (before the Hubbard splitting). This $Q_3=0$ boundary just separates $\Phi_1$ and $\Phi_3$, although both of them are $J_{\rm eff}=3/2$ states.

This transition can be also evidenced in the magnetic moments, as shown in Fig.~\ref{F4}(d). Although the net magnetic moment is always close to zero in the whole range, the spin magnetic moment and orbital magnetic moment change from $\sim0.3$ $\mu_{\rm B}$/Ta to $\sim0.7$ $\mu_{\rm B}$/Ta.

All above results demonstrated that the $J_{\rm eff}$ states in $4d^1$ and $5d^1$ systems could be efficiently tuned by strain, via the tetragonal distortion.

\section{Summary}
Our theoretical studies, based on both the atomic orbital model and first-principles calculations, clarified the divergent results regarding the $J_{\rm eff}=3/2$ states in $4d^1$ and $5d^1$ systems. Different from the previous argument which relied on the difference of SOC intensity, our study highlighted the key role of $Q_3$ mode distortion. With different signs of $Q_3$ mode and SOC intensities, four typical hexachlorides, K$_2$NbCl$_6$, Rb$_2$NbCl$_6$, K$_2$TaCl$_6$, and  Rb$_2$TaCl$_6$, own three types of electronic states: $J_{\rm eff}=3/2$ state $\Phi_1$ (or $\Phi_2$) for both K$_2$NbCl$_6$ and K$_2$TaCl$_6$; $J_{\rm eff}=3/2$ state $\Phi_3$ (or $\Phi_4$) for Rb$_2$TaCl$_6$; but $S=1/2$ state $d_{xy}$ for Rb$_2$NbCl$_6$. In addition, our work also proposed an efficient route to control these electronic states in $4d^1/5d^1$ systems. By tuning the tetragonal distortion through epitaxial strain, quantum phase transitions between these three states can be achieved. Following experiments are encouraged to manipulate these $J_{\rm eff}=3/2$ quantum states in this way. More theoretical investigations on other distortions beyond the simplest $Q_3$ mode are also encouraged, which may lead to more emergent physical phenomena in these strong SOC quantum materials.

\begin{acknowledgments}
This work was supported by the National Natural Science Foundation of China (Grant Nos. 11804168 and 11834002), the Natural Science Foundation of Jiangsu Province (Grant No. BK20180736), and NUPTSF (Grant No. NY219026).
\end{acknowledgments}

\bibliographystyle{apsrev4-1}
\bibliography{ref}
\end{document}